\documentclass[namedreferences]{solarphysics}
\usepackage[optionalrh]{spr-sola-addons} 
\usepackage{graphicx}                    
\usepackage{color}                       
\usepackage{ulem}                      
\usepackage{url}                         
\usepackage{rotating}

\begin{document}

\begin{article}

\begin{opening}

\title{Fitting and Reconstruction of Thirteen Simple Coronal Mass Ejections }
\author{Nada~\surname{Al-Haddad}$^{1}$, Teresa~\surname{Nieves-Chinchilla}$^{1,2}$, Neel~P.~\surname{Savani}$^{2,3}$, No{\'e}~\surname{Lugaz} $^{4}$, Ilia~I.~\surname{Roussev}$^{5}$}

\runningauthor{Al-Haddad et al.}
\runningtitle{Fitting and Reconstruction of Thirteen Well-Observed CMEs}

\institute{
$^{1}${IACS-Catholic University of America, Washington, DC, USA.}
$^{2}${NASA Goddard Space Flight Center, Greenbelt, MD, USA.}
$^{3}${University of Maryland Baltimore County, Baltimore, MD, USA.}
$^{4}${Space Science Center and Department of Physics, University of New Hampshire, Durham, NH, USA}
 $^{5}${Center for mathematical Plasma Astrophysics, Katholieke Universiteit Leuven, Leuven, Belgium}}

\begin{abstract}
Coronal mass ejections (CMEs) are the main drivers of geomagnetic disturbances, but the effects of their interaction with Earth's magnetic field depend on their magnetic configuration and orientation. Fitting and reconstruction techniques have been developed to determine the important geometrical and physical CME properties, such as the orientation of the axis, the CME size, and the magnetic fluxes, among others. In many instances, there is disagreement between such different methods but also between fitting from {\it in situ} measurements and reconstruction based on remote imaging. This could be due to the geometrical or physical assumptions of the models, but also to the fact that the magnetic field inside CMEs is only measured at one point in space as the CME passes over a spacecraft. Here, we compare three methods based on different assumptions for measurements of thirteen CMEs by the  {\it Wind} spacecraft from 1997 to 2015. 
 These CMEs are selected from the interplanetary coronal mass ejections catalog on {\url{https://wind.nasa.gov/ICMEindex.php} }due to their simplicity in terms of 1) small expansion speed throughout the CME and 2) little asymmetry in the magnetic field profile. This makes these thirteen events ideal candidates to compare codes that do not include expansion nor distortion. We find that, for these simple events, the codes are in relatively good agreement in terms of the CME axis orientation for six out of the 13 events. Using the Grad-Shafranov technique, we can determine the shape of the cross-section, which is assumed to be circular for the other two models, a force-free fitting and a circular-cylindrical non-force-free fitting. Five of the events are found to have a clear circular cross-section, even when this is not a pre-condition of the reconstruction. We make an initial attempt at evaluating the adequacy of the different assumptions for these simple CMEs. 
The conclusion of this work strongly suggests that attempts at reconciling {\it in situ} and remote-sensing views of CMEs must take in consideration the compatibility of the different models with specific CME structures to better reproduce flux ropes.
\end{abstract}
\end{opening}

\section{Introduction}
Coronal mass ejections (CMEs) have been measured {\it in situ} and observed remotely for several decades \cite{Gold:1962,Tousey:1973,Burlaga:1981}. A recent review and historical perspective on the study of CMEs can be found in \inlinecite{Gopalswamy:2016}. Following numerous models and studies ({\it e.g.}, see \opencite{Gold:1960}), \inlinecite{Burlaga:1981}, \inlinecite{Klein:1982}, \inlinecite{Goldstein:1983}, \inlinecite{Zhang:1988} and \inlinecite{Lepping:1990} developed the current paradigm of the magnetic cloud (MC) CME, composed of helically closed field lines which can be fitted with force-free models. It is understood that not all CMEs are MCs, the proportion of MC {\it vs}.\ non-MC CMEs varies with the solar cycle \cite{Richardson:2004b,Richardson:2010,Jian:2006}. In this article, we use the term CMEs to describe ejections measured {\it in situ} as well as observed remotely. For {\it in situ} measurements, we employ the term CME to refer to the entire observed structure, including the shock and the sheath, when one is present, as well as the magnetic obstacle (MO, see \opencite{Nieves:2018}), which may or may not be a MC. 

The investigation of CMEs has progressed in the past decade, due to the availability of multi-spacecraft measurements combining measurements from the {\it Solar-Terrestrial Relations Observatory} (STEREO), the {\it Advance Composition Explorer} (ACE) and {\it Wind} as well as measurements from planetary missions \cite{Kilpua:2009a,Moestl:2009d,Farrugia:2011,Nieves:2012,Winslow:2016}, remote heliospheric observations \cite{Bisi:2008,Davies:2009,Savani:2009,Nieves:2013}, and numerical simulations \cite{Lynch:2008,Manchester:2008,Jacobs:2009,Alhaddad:2011,Lugaz:2011c,Roussev:2012,Kliem:2013}. Through these, the complexity of the formation, propagation and interaction of CMEs and the potential intricacy of their internal structures have been pointed out. In fact, more than 40 years after the first remote detection of a CME, the space science community is still debating the exact cause and morphology of CMEs. Due to this, several models have been developed to fit and reconstruct {\it in situ} measurements of CMEs, based on different assumptions. Because spacecraft only measure a time series at one point (which can be converted into a 1D cut along the direction of crossing) and CMEs are inherently three dimensional (3D), any fitting and reconstruction method must be based on some assumption of symmetry. \inlinecite{Riley:2004} used a numerical simulation, where the CME morphology and structure are known to test different fitting and reconstruction techniques for a CME crossing close to its ``nose'' and one at a high impact parameter, far from the center of the magnetic structure. There, the authors found that, for large impact parameters, different techniques can result in very different fitted characteristics of the CMEs. The authors also raised the issue of the choice of boundaries, which can differ significantly from one fitting to another, and likely contributes to the difference found between different techniques. In \inlinecite{AlHaddad:2013}, we compared several different fitting and reconstruction techniques for observed (rather than simulated) events, as part of a coordinated data analysis workshop (CDAW), and confirmed that there can be significant differences between techniques but also that, by choosing the same boundaries, it is common that two or more techniques are in close agreement (see also work by \opencite{Ruffenach:2012} and references therein). 

One of the main questions underlying the choice of a fitting and reconstruction model, beside the choice of the boundaries, is to determine which approximation is the most adapted to a particular event or to all CMEs in general. Following \inlinecite{Zhang:1988} and \inlinecite{Lepping:1990}, force-free fitting with a Lundquist solution is one of the most commonly used methods. It is assumed that the field is force-free with a geometrically simple and circular cross section, assuming a straight axis, and an axisymmetric configuration. On the other hand, the Grad-Shafranov reconstruction \cite{Hu:2001,Moestl:2009,Isavnin:2011} assumes that magnetic field inside a CME is in magneto-hydrostatic (MHS) equilibrium and reconstruct the field without a pre-imposed cross-section.  However, the Grad-Shafranov technique still relies on the assumption of invariance along a straight axis, {\it i.e.} it assumes a translational symmetry.  In \inlinecite{Alhaddad:2011}, it was pointed out that this assumption of translational symmetry may affect the realism of the reconstruction for true 3D magnetic configurations. The model of \inlinecite{Hidalgo:2000} and \inlinecite{Hidalgo:2002b} has recently been generalized in \inlinecite{Nieves:2016}. This model assumes neither force-free nor MHS equilibrium conditions, but solves the Maxwell equations for a given current. In its current version, the model assumes a circular cross-section and axial symmetry. 

Here, we compare these different models for thirteen well-observed CMEs that occurred between 1995 and 2015. Our initial hypothesis is that studies such as that performed by \inlinecite{Riley:2004} can give us some information about the bias and performance of fitting and reconstruction techniques but are limited by the necessary simplifications in numerical models (for example, assuming the plasma can be treated as fluid in MHD, assumptions related to the CME initiation methods, to the solar wind model, {\it etc.}). Here, we attempt at comparing the different techniques in terms of which one is the most adapted for selected measured CMEs.  
In Section~\ref{sec:Data}, we discuss our data selection and give references for the codes and fitting procedures used in this study. In Section~\ref{sec:Results}, we present fitting results for these thirteen CMEs and compare the performance of the codes for each CMEs. We discuss our findings and conclude in Section~\ref{sec:Conclusions}.

\section{Data and Techniques} \label{sec:Data}
Our analysis relies on measurements made by the {\it Wind} spacecraft since 1995, specifically by the {\it Magnetic Field Investigation} (MFI, \opencite{Lepping:1995}) instrument and the {\it Solar Wind Experiment} (SWE) and {\it Three-Dimensional Plasma and Energetic Particle Investigation} (3DP) \cite{Ogilvie:1995,Lin:1995} plasma instruments. We start {\it ab initio} from {\it Wind} measurements, including the magnetic field strength, components, the proton density, temperature, velocity and $\beta$ to create a new catalog of CMEs measured {\it in situ} near L1. In this same issue, \inlinecite{Nieves:2018} explore {\it  in situ} signatures of CMEs to  better understand  the internal structure of their magnetic field. Through that study, this comprehensive database of CME measurements by {\it Wind} has been made publicly available at {\url{https://wind.nasa.gov/ICMEindex.php} } and henceforth it will be referred as the catalog. 
The goal of creating a new catalog instead of starting from existing ones, such as that of \inlinecite{Richardson:2010}, \inlinecite{Jian:2006}, or \inlinecite{Lepping:2006} was to include cases with clear signatures of an organized magnetic structure, and exclude cases corresponding to complex ejecta \cite{Burlaga:2002}. Rather than relying on the strict definition of a magnetic cloud following \inlinecite{Burlaga:1981}, this catalog focuses on magnetic obstacles (MOs) that are characterized by low proton plasma, $\beta_p$, low proton temperature, and an organized magnetic topology. Such events have not necessarily all been captured in these previous lists.

We identify 337 MOs that meet these criteria, many of which are small transients \cite{Yu:2014} or events with relatively weak increase in the magnetic field strength (see \opencite{Nieves:2018} for the full description of the catalog). From this initial list, we further check the magnetic hodograms, identifying events for which there is a clear rotation in the hodogram, and a clear single flux rope. We also only keep events with a duration of at least 12 hours and an increase of the magnetic field of at least 50\% with respect to the background. Lastly, we calculate the expansion speed and DiP (Distortion Parameter, \opencite{Nieves:2018}) for each of these events. 

The DiP is a measure of the symmetry of the magnetic field profile. The parameter is derived from the temporal average of the magnetic field,  $\langle B \rangle$. The DiP value is the fraction of the total duration where 50\% of the total $\langle B \rangle$  is accumulated. Therefore, by definition DiP varies from 0 to 1. Values of DiP less than 0.5 imply compression at the leading edge and values greater than 0.5 imply compression at the back of the structure. A DiP value of 0.5 implies that the magnetic field strength is balanced between the front half and back half of the CME, {\it i.\ e}.\ that there is no significant amount of front or back compression. We select 13 cases spread throughout the 21 years of the sample and for which the DiP is close to 0.5 (between 0.44 and 0.57) and the expansion speed is low, close to 0~km\,s$^{-1}$ (between $-15$ and 29~km\,s$^{-1}$). The expansion speed is derived from a simple linear fit of the solar wind bulk velocity in the MO interval. As such, these 13 events shall not be thought of as typical cases but as cases as close as possible to an ideal non-expanding single flux rope measured near 1~AU. 

These 13 events are summarized in Table~1. As can be seen, these events are typically slow with an average speed of 380~km\,s$^{-1}$, ranging between 300 and 500~km\,s$^{-1}$ and with an expansion speed that is typically about 5\% of the CME propagation speed (as compared to 15--20\% for typical CMEs at 1~AU, see {\it e.g.} \opencite{Farrugia:1993}). The average duration of these events as given by the Grad-Shafranov reconstruction code is about 19 hours, which is typical for magnetic clouds at 1 AU \cite{Richardson:2010}. Most of these events are in fact magnetic clouds, as defined by \inlinecite{Lepping:1990}.

\begin{table}[ht]
\begin{center}
\centering
\begin{tabular}{|ccccc|c|c|c|}
\hline
\multicolumn{5}{|c|}{Event} & V & DiP & V$_\mathrm{exp}$ \\
\hline
Year & DOY & MM/DD & HH:MM & $\Delta$t (h)  & (km\,s$^{-1}$) & & (km\,s$^{-1}$)\\
\hline
1997 & 10 & 01/10 & 06:18 & 16.7 & 437 & 0.50 & 29\\ 
1998 & 232 & 08/19 & 15:58 & 22.7 & 317 & 0.44 & 27\\ 
2000 & 183 & 07/01 & 07:48 & 21.5 & 413 & 0.49 & $-13$\\
2001 & 111 & 04/21 & 23:58 & 21.5 & 357 & 0.44 & 29\\
2002 & 273 & 09/29 & 23:48 & 19.2 & 380 & 0.46 & $-15$\\
2008 & 143 & 05/23 & 23:18 & 10.7 & 499 & 0.51 & 21\\
2009 & 273 & 09/30 & 07:38 & 9.2 & 346 & 0.51 & 12\\ 
2012 & 137 & 05/16 & 17:18 & 22.7 & 374 & 0.57 & $-7.0$ \\ 
2012 & 317 & 11/12 &10:18 & 15.7 & 381 & 0.48 & 1.0\\ 
2013 & 179& 06/27 & 01:58 & 30.7 & 391 & 0.51 & 7.0\\ 
2013 & 359 & 12/24 &  04:58 & 12.2 & 299 & 0.50 & 23\\ 
2014 & 101 & 04/11 & 19:18 & 23.5 & 350 & 0.55 & 24\\ 
2015 & 126 & 05/06 & 16:08 & 19.8 & 422 & 0.46 & 18\\
\hline 
\end{tabular}
\caption{List of 13 CMEs used in this study, including start time and duration from the Grad-Shafranov reconstruction, average speed, DiP parameter and expansion speed. DOY corresponds to the day of the year in {\it Wind} data, and the date is shown as month and day (MM/DD).}
\end{center}
\end{table}

The goal of this selection is to identify events for which flux rope models with a circular cross-section and force-free configuration might be the most appropriate. Although the inclusion of radial expansion in force-free models has been described and implemented over 20 years ago \cite{Farrugia:1993,Marubashi:2007,Nakwacki:2008}, most models used to fit {\it in situ} measurements to determine the CME properties do not include expansion, and this is the case for all models in this study. We note that a non-expanding CME may be in itself highly unusual and reflect an unusual interaction with the solar wind \cite{Gulisano:2010}; however, our focus is not on the origin of the symmetry and small expansion speeds, but rather on the validity of codes to fit and reconstruct such ``ideal'' CMEs. 
For each of the 13 selected events, we perform fits and reconstruction with the force-free (FF) fitting of \inlinecite{Savani:2011b}, the circular-cylindrical (CCS) flux rope model of \inlinecite{Nieves:2016} and with the Grad-Shafranov (GS) reconstruction code \cite{Hu:2001,Moestl:2008}. These three models do not include expansion in their formalism and therefore are appropriate for CME with minimal radial expansion. When performing the reconstruction or fitting, the data, initially at about 1-min resolution, is binned into bins of duration between 0.5 and 1.5 hours. For the Grad-Shafranov reconstruction, we use 18--25 bins. For the fitting methods, we use 17--61 bins, which correspond to either 30 minutes {\it per} bin or 1 hour  {\it per} bin. In order to compare the different results quantitatively, we use the reduced version of the chi-squared parameter, $\chi_\mathrm{red}$, which takes into consideration the number of free parameters of the models as well as the number of datapoints (see for example \opencite{Lepping:2006}). $\chi_\mathrm{red}$ is defined as: $\chi^2_\mathrm{red} = \frac{\chi^2}{N-f}$, or
$$\chi_\mathrm{red} = \sqrt{\frac{\sum_{i = 0}^N (y_{mi} - y_{di})^2}{N-f}},$$
where $N$ is the number of datapoints, $f$ the number of free parameters of the model, $y_{mi}$ the value of the fitted quantities of the model (see Section~2) at the point $i$ and $y_{di}$ the value of the data at the point $i$. 

The selection of the boundaries is a frequent source of disagreement in the results of fitting and reconstruction methods, as pointed out in \inlinecite{Riley:2004}, \inlinecite{AlHaddad:2013}, \inlinecite{Ruffenach:2012}, among others. In order to compare the codes without the influence of the boundary selection, we use the boundaries found during the Grad-Shafranov reconstruction for the two fitting codes. These boundaries are selected in the process of fitting the total transverse pressure, $P_t$, as a function of the magnetic potential, $A$, by enforcing that the inbound and outbound crossings through the flux rope start at a similar value of $A$ \cite{Hu:2004}. In \inlinecite{AlHaddad:2013}, we found that using the Grad-Shafranov boundaries for the other codes makes the results of different codes more consistent between each other.

The force-free fitting with a circular cross-section flux rope is one of the simplest fitting techniques, initially developed by \inlinecite{Lepping:1990}, following \inlinecite{Marubashi:1986} and \inlinecite{Burlaga:1988}. In the version used in this study, the flux rope orientation is obtained using a minimum variance analysis \cite{Sonnerup:1998}. The fitting is done to minimize $\chi_\mathrm{red}$ to determine the strength of the magnetic field and impact parameter (closest approach distance to the CME axis). The $\chi_\mathrm{red}$ is calculated using the three components of the magnetic field. The radius of the flux rope is obtained directly without fitting, once the orientation is found using minimum variance analysis, given the boundaries and velocity measurements. Therefore, the force-free fitting technique employed here has two free parameters, the magnetic field and impact parameter.

\begin{figure*}[tb]
\centering
\includegraphics[width=1.02\hsize]{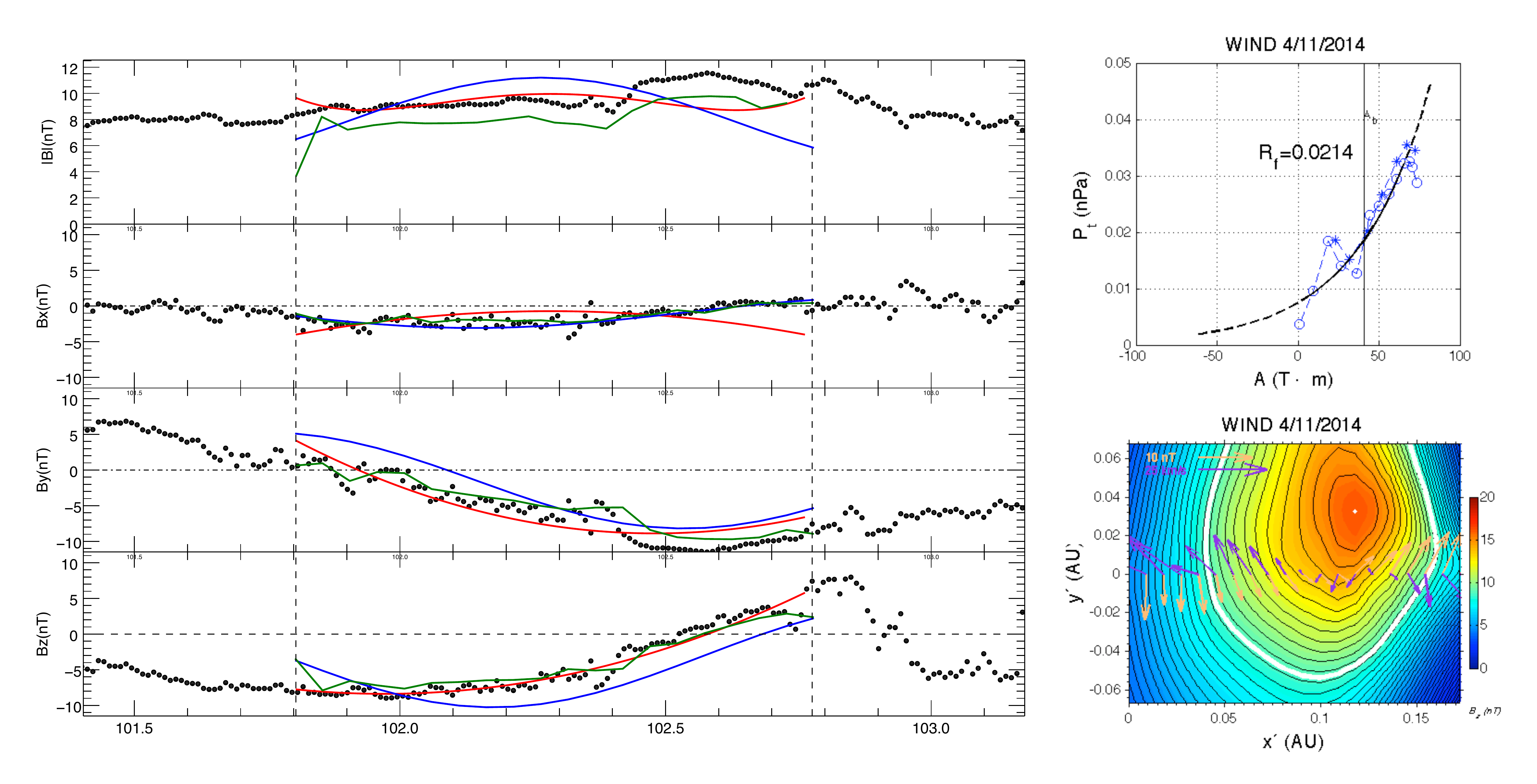}
\caption{CME on 11 April 2014 (2014/101, year DOY): Example of a CME reconstructed with a nearly circular cross-section by the Grad-Shafranov technique, but for which the axis orientation from the three models deviates significantly from each other (GS and CCS are within 52$^\circ$). The left panel show the force-free fitting (blue), the circular-cylindrical non-force-free fitting (red), and the Grad-Shafranov construction (green). In this panel the horizontal axis corresponds to the DOY. The right panels show the Grad-Shafranov fitting of the total transverse pressure by the magnetic potential (top) and the Grad-Shafranov reconstruction (bottom). All angles and components are in the Geocentric Solar Ecliptic (GSE) coordinate system.} 
\label{fig:circular}
\end{figure*}

The circular-cylindrical flux rope model of \inlinecite{Nieves:2016} builds upon the non-force-free models of \inlinecite{Hidalgo:2000} and \inlinecite{Hidalgo:2002}. In it, the Maxwell equations are solved for a given current density profile and orientation. Current and impact parameter of the flux rope are obtained by minimizing $\chi_\mathrm{red}$, which is calculated for the three components of the magnetic field and the magnitude of the magnetic field. The model has five free parameters, the orientation of the flux rope axis (2 angles), the strength of the axial current, $\alpha$ which is related to the ratio of poloidal to axial currents, and the impact parameter (see \inlinecite{Nieves:2016} for more details). 

The Grad-Shafranov reconstruction technique differs from the two previously described fitting techniques as well as most other techniques in the sense that: 1) the selection of boundaries is a step occurring during the reconstruction, not done before the fitting, 2) the shape of the cross-section is not pre-determined, and 3) it is based on a polynomial and exponential fit of the satellite data followed by an integration away from the data to reconstruct the flux rope. As such, the Grad-Shafranov reconstruction is not a fitting technique and errors in the reconstruction are only based in errors or limitations in the fitting of the total transverse pressure (apart from the limitation of the  applicability of the MHS approximation for the CME at the core of the method). $\chi_\mathrm{red}$ corresponds to the fitting of the total transverse pressure as a function of the magnetic potential, $A$ (see \opencite{Moestl:2008}, for example, for details). For the Grad-Shafranov technique, we adjust the boundaries and the way the exponential function is combined with the polynomial to fit the total transverse pressure (two parameters). As such, this reconstruction technique has four free parameters (two for boundaries and two for the fitting).  

We also compare our list of 13 events with the list of magnetic clouds of \inlinecite{Lepping:2006} and the list of interplanetary CMEs of \inlinecite{Richardson:2010}. The list of magnetic clouds is updated online to 2012, whereas the list of interplanetary CMEs is updated more frequently to about six months before the current date. All of our events are in the CME list except for the one in May 2008. All our events from 1996 to 2012 are magnetic clouds, except for that  event. This shows that magnetic obstacles, following our definition, with low DiP and low expansion speed are almost all magnetic clouds. 

\section{Results} \label{sec:Results}

We perform the fittings and reconstruction of these 13 CMEs using the three methods discussed above. Our results are presented in Table~2, where all angles are given in the GSE coordinate system. We compare the reduced $\chi_\mathrm{red}$ for each fit and model. Note that $\chi_\mathrm{red}$ displayed in the Grad-Shafranov column differs from the residue usually plotted for Grad-Shafranov reconstruction. The reduced $\chi_\mathrm{red}$ takes into account on how many datapoints the fit is performed and how many free parameters the model has. Using $\chi_\mathrm{red}$ allows, therefore, to take into consideration the fact that the fitting is done on different numbers of datapoints, due to different binning, and also that the different models have different numbers of free parameters. 
$\chi_\mathrm{red}$ here, is not a dimensionless number. The Grad-Shafranov reconstruction fits the total transverse pressure to a polynomial function of the magnetic potential,  while, the CCS and the FF fit the magnetic field. Therefore, the different $\chi_\mathrm{red}$ are not directly comparable quantities.

Note that for the Grad-Shafranov reconstruction the error increases away from the satellite path because of the integration errors (see \opencite{Hu:2001} for details); therefore, $\chi_\mathrm{red}$ does not quantify the accuracy of the reconstruction away from the satellite path. We find that, for nearly all cases the CCS fitting returns lower $\chi_\mathrm{red}$ values than the force-free fitting using minimum variance. This  indicates that, even for relatively simple events, a non-force-free method, such as CCS, might be preferable as compared to a force-free fitting method.

\begin{table}[ht]
\footnotesize
\begin{center}
\centering
\tabcolsep=0.11cm
\begin{tabular}{|cc|ccc|cccc|ccc|}
\hline
\multicolumn{2}{|c|}{Event} & \multicolumn{3}{|c|}{CCS} & \multicolumn{4}{|c|}{GS} & \multicolumn{3}{|c|}{FF}\\
\hline
Year & DOY & $\theta(^\circ)$ & $\Phi(^\circ)$ & $\chi_\mathrm{red}$ (nT) & $\theta(^\circ)$ & $\Phi(^\circ)$ & $\chi_\mathrm{red}$ (nPa) & b/a &$\theta(^\circ)$ & $\Phi(^\circ)$ & $\chi_\mathrm{red}$ (nT)\\
\hline
1997 & 10 & 232& $-21$ & 0.028 & 214 & $-16$ & 0.017 & 0.5 & 231 & $-27$ & 0.038 \\
1998 & 232 & 264 & 17 & 0.044 & 265 & 13 & 0.017 & 0.6 & 294 & 28 & 0.050\\
2000 & 183 & 359 & 21 & 0.045 & 2 & 41.5 & 0.023 & 0.6 & 357 & 48 & 0.11\\
2001 & 111 & 285 & $-42$ & 0.025 & 302 & $-35$ & 0.017 & 0.9 & 272 & $-48$ & 0.062\\
2002 & 273 & 109 & $-12$ & 0.035 & 146 & $-12$ & 0.015 & 1  & 114 & 18 & 0.11\\
2008 & 143 & 24 & 47 & 0.033 & 20 & 12 &  0.015 & 0.6  & 306 & 68 & 0.075\\
2009 & 273 & 284 & 36 & 0.036 & 323 & 14 &  0.017 &  0.6 & 86 & $-1$ & 0.101\\
2012 & 137 & 7 & $-1$ & 0.021 & 105 & $-13$ & 0.018 & 0.7  & 285 & 10 & 0.075\\
2012 & 318 & 283 & $-55$ & 0.023 & 268 & 40 &  0.023 & 0.6 & 301 & $-30$ & 0.050\\
2013 & 179 & 283 & $-55$ & 0.018 & 330 & $-66$ & 0.018 & 0.8 & 240 & $-62$ & 0.050\\
2013 & 359 & 26 & $-65$ & 0.063 & 53 & $-50$ & 0.019 & 0.5 & 44 & $-74$ & 0.043\\
2014 & 101 & 250 & $-38$ & 0.021 & 236 & $-27$ & 0.021 & 0.9 & 116 & 51 & 0.10\\
2015 & 126 & 133 & $43$ & 0.041 & 158 & 49 & 0.021 & Cx & 294 & 28 & 0.053\\
\hline 
\end{tabular}
\caption{Axis direction given in terms of longitude, $\theta$, and latitude, $\Phi$, and $\chi_\mathrm{red}$, for the circular-cylindrical fitting model (CCS), the Grad-Shafranov model (GS), and the force-free fitting model (FF). For GS, we also list the aspect ratio of the cross-section ($b/a$). Cx refers to complex cross-sections.}
\end{center}
\end{table}

We also calculate the angle between the reconstructed axis for the same event but for different models. In six cases (1997/10, 1998/232, 2000/183, 2001/111, 2008/143 and 2013/179) all models find axis directions within 46$^\circ$ (we include one case for which two models are 46$^\circ$ apart), in two additional cases (2002/273 and 2012/137) at least two models return a MO axis direction within 45$^\circ$; in another five cases all models differ by more than 46$^\circ$ from one another. This actually corresponds to a better agreement between the different models than found previously \cite{AlHaddad:2013,Riley:2004}.

\begin{figure*}[tb]
\centering
\includegraphics[width=1.02\hsize]{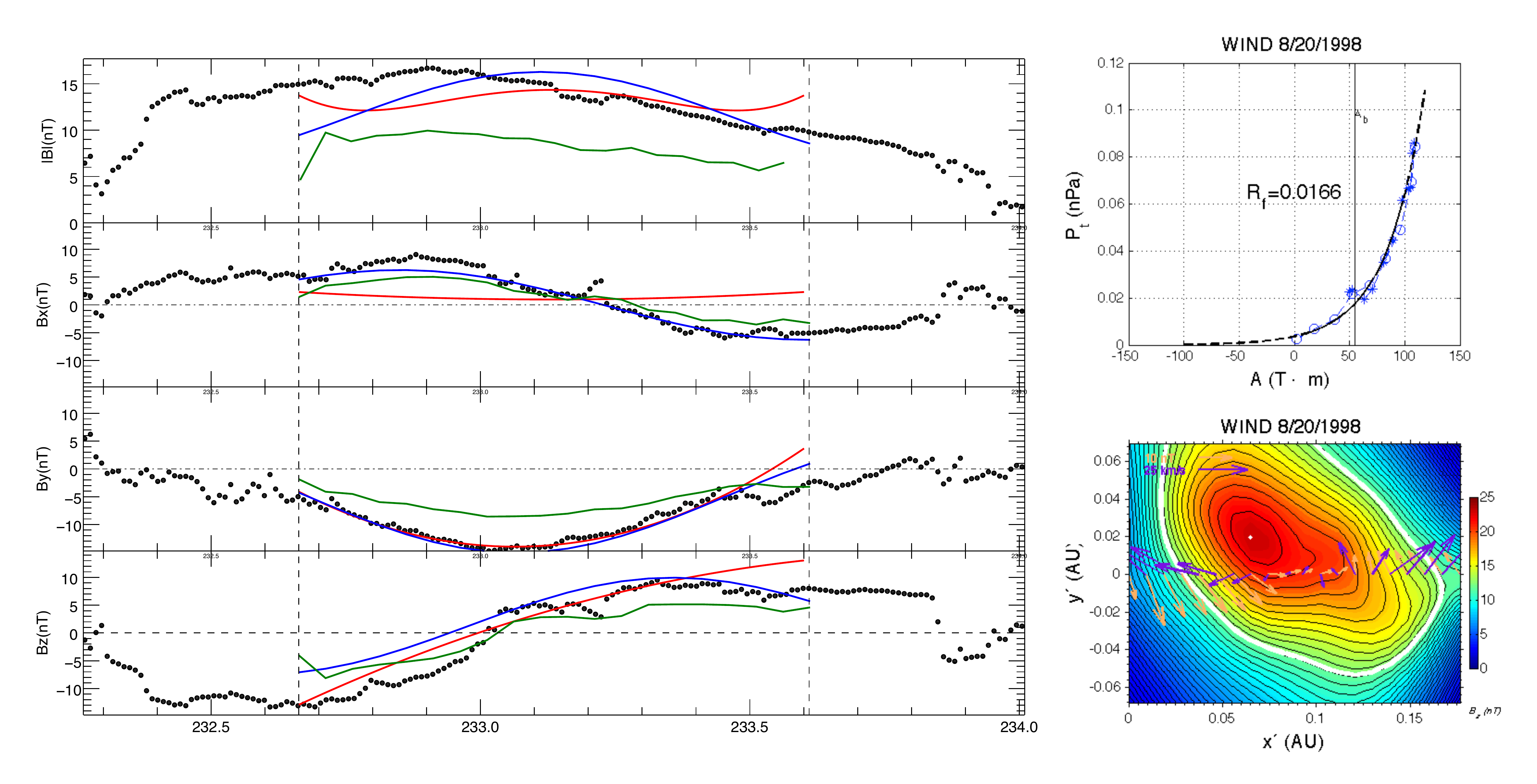}
\caption{CME on 19 August 1998 (1998/232, year/DOY): Example of a CME reconstructed with an elliptical cross-section by Grad-Shafranov but for which the axis orientation from the three models is within 30$^\circ$ from each other (GS and CCS are within 4$^\circ$). The left panel shows the force-free fitting (blue), the circular-cylindrical non-force-free fitting (red), and the Grad-Shafranov construction (green). In this panel the horizontal axis corresponds to the DOY. The right panels show the Grad-Shafranov fitting of the total transverse pressure by the magnetic potential (top) and the Grad-Shafranov reconstruction (bottom).} 
\label{fig:elliptical}
\end{figure*}


We visually inspect the cross-section of the Grad-Shafranov reconstructions for each event and characterize them into three main categories: 1) circular, when the aspect ratio of the ellipse is between 0.8 and 1, 2) elliptical when the aspect ratio is less than 0.8, 3) complex when the shape is more complicated than an ellipse or circle. We find that four cases can be characterized as having a circular cross-section (see Figure~\ref{fig:circular} for an example), eight as having an elliptical cross-section (see Figure~\ref{fig:elliptical} for an example), and one as having a complex cross-section (typically with two local maxima of the magnetic field in addition to an elliptical or more complex shape). There seems to be no one-to-one relationship between the shape of the cross-section and how well the three models agree on the axis orientation. It should be noted, however, that the  complex event has strong disagreement between the three models, whereas three of the four circular cross-section cases have good agreement between all three models. In addition, the smallest improvements obtained by using the CCS fitting rather than the force-free fitting, as measured by the decrease in $\chi_\mathrm{red}$, occur for those events that the Grad-Shafranov reconstruction shows an elliptical or complex cross-section.

\section{Discussion and Conclusions} \label{sec:Conclusions}

This is an initial attempt at determining how well fitting models and reconstruction techniques work for various cases of simple CMEs. We focus on simple flux ropes with minimal expansion, distortion and asymmetry in the magnetic field profile as candidates to test  models with simple geometry, a force-free model, a non-force-free model, and a reconstruction technique. We find a relatively good agreement between the different fitting results for six out of the thirteen events considered. In particular, for seven out of the thirteen selected events, the two non-force-free codes find a CME axes within $45^\circ$ of each other (in six of these cases, the orientation is also within 45$^\circ$ of that given by the minimum variance analysis). We also determine the shape of the cross-section of these CMEs based on the Grad-Shafranov reconstruction, and find that not all simple events correspond to a CME with a circular cross-section. Our results suggest that, for those events where the Grad-Shafranov reconstruction returns a circular-cross-section, the other models that assume a circular-cylindrical geometry are in relatively good agreement (for three of the four such cases, all three models find a CME axis within 45$^\circ$). This result suggests that the Grad-Shafranov reconstruction could be a good tool to discriminate models based on different geometries.

Future work will continue in three main directions: 1- this study will be extended to more events, in particular for some with a clear asymmetry in the magnetic field profile (DiP $\ne$ 0.5) and/or expansion speed; 2- we will identify the source region of the CME events and their coronal counterparts (especially for events since 2008 when the stereoscopic viewpoints of STEREO are available) and determine the orientation of the CME in the corona and inner heliosphere, as was done, for example in \inlinecite{Nieves:2012}; 3- non-force-free models with elliptical cross-sections should be further developed and compared to Grad-Shafranov reconstructions for events where the latter method indicates an elliptical cross-section. It is often assumed that magnetic ejecta start with a circular cross-section, and that expansion and interaction with the solar wind result in the cross-section to become elliptical. This is for example the case in most numerical simulations \cite{Riley:1997,Vandas:2002,Manchester:2004,Owens:2006}. However, since measurements of  the expansion speed are performed at 1~AU, they do not necessarily give information on the past behavior of the CME (regarding if and how it expanded) and this may explain why we find a significant number of elliptical cross-sections, even for events specifically selected for the small magnitude of the expansion speed at 1~AU. Another reason might be that CMEs start with an elliptical cross-section already in the corona, although some evidence from EUV and coronagraph images point to circular cross-sections \cite{Vourlidas:2017}.

\vspace{0.3cm}

\noindent {\bf Disclosure of potential conflicts of interests:} The authors declare that they have no conflicts of interest.

 \begin{acks}
The authors would like to thank the reviewer for their useful comments that helped improving this manuscript.
We acknowledge the use of {\it Wind} data provided by the
magnetometer and the solar wind experiment teams at GSFC.
This research was performed with funding from AGS-1433086, AGS-1433213, and AGS-1460179.
 \end{acks}

\bibliographystyle{spr-mp-sola-cnd}

\end{article}

\end{document}